\documentclass[preprint,review,12pt]{elsarticle}

\usepackage{amssymb}

\usepackage{graphicx}
\usepackage{amsmath}
\usepackage{hyperref}
\usepackage{xspace}

\usepackage{color}
\definecolor{darkblue}{rgb}{0,0,0.5}
\definecolor{darkred}{rgb}{0.5,0,0}
\definecolor{darkgreen}{rgb}{0,0.5,0}

\newcommand{\be}{\begin{equation}}
\newcommand{\ee}{\end{equation}}

\usepackage{cleveref}
\usepackage{float}
\usepackage{tikz}
\usetikzlibrary{arrows,arrows.meta,hobby,patterns}
\usepackage{morefloats}
\usepackage{mathrsfs}
\usepackage{listings}
\definecolor{droplet1}{rgb}{0.878,0.953,0.859}
\definecolor{droplet2}{rgb}{0.659,0.867,0.710}
\definecolor{droplet3}{rgb}{0.263,0.635,0.792}

\newcounter{bla}

\journal{Computer Physics Communications}

\begin{document}

\begin{frontmatter}



\title{Efficient Modeling of Particle Transport through Aerosols in GEANT4}


\author[a]{Nathaniel J.L. MacFadden}
\author[a]{Ara N. Knaian\corref{author}}

\cortext[author] {Corresponding author.\\\textit{E-mail address:} ara@nklabs.com}
\address[a]{NK Labs, Cambridge, MA 02139, USA}

\begin{abstract}
We present a geometry class for efficiently simulating particle transport through aerosols in GEANT4. It is demonstrated that aerosol granularity can strongly affect this transport and thus a generic aerosol model must respect this granularity, which the presented class achieves by modelling the aerosol as a collection of droplets. For large aerosols, this class is orders of magnitude quicker and less memory intensive than standard granularity-respecting methods to model aerosols in GEANT4. These gains are allowed by simpler voxelization optimization, by only populating droplets relevant to the transport, and by using droplet geometry to consider fewer droplets per calculation. The presented class allows differing aerosol bulk/droplet shape, droplets with structure, and spatially varying droplet position/rotation.
\end{abstract}

\begin{keyword}
aerosol; particle transport; GEANT4

\end{keyword}

\end{frontmatter}


\section{Introduction}
GEANT4 is a C++ based open-source toolkit for simulating the passage of particles through matter\cite{doi:10.1016/S0168-9002(03)01368-8}. Often used for high energy physics and accelerator design, GEANT4 also has a range of scientific/engineering applications such as design of radiation shielding, medical equipment, and space hardware\cite{doi:10.1016/j.nuclphysbps.2004.10.083,doi:10.1016/j.nima.2016.06.125}. The transport of particles through aerosols is important both in traditional application such as particle detectors\cite{doi:10.1038/srep41699, doi:10.1016/j.jenvrad.2016.03.025} and in many alternate applications such as environmental studies\cite{isbn:0521807670, doi:10.1007/978-0-387-87664-1_10, doi:10.5772/50248} and astronomical studies\cite{doi:10.3847/0004-637x/818/2/164, doi:10.1088/0004-637x/737/2/105, doi:10.1086/590482, doi:10.3847/1538-4357/ab1b20}, but this transport is complicated\cite{doi:10.1016/j.jaerosci.2018.12.010} since, for some aerosols, their granularity is important\cite{doi:10.3847/0004-637x/818/2/164}.

The simplest aerosol model, a single-volume mixture of the background and droplet materials in the correct mass fractions, is sufficient when this granularity is unimportant, but we experimentally find this only to be true when droplets are small. The failure of this approximation can be understood because a single volume model implicitly posits equal interaction probability throughout the aerosol (uniform distribution of droplet material) but, when aerosols consist of large droplets, there is a heightened probability for interactions in quick succession.

The granularity is most directly respected by modelling the aerosol as a collection of (non-intersecting) droplets, which can be done with GEANT4's `parameterised solid' geometry. A parameterised solid consists of one or more sub-objects that are repeatedly placed in the world according a parameterisation function that allows differences in position, size, rotation, material, and more between copies\cite{bookForApplicationDevelopers}. While such solids are maximally flexible, they experience computational difficulties when the sub-object is populated millions or billions of times, as in the case of many aerosols. To allow for such aerosols with many large droplets to be simulated in GEANT4, we present the `fastAerosol' geometry class which models the aerosol as a collection of droplets but with three following differences from parameterised solids.

First, while both fastAerosol and parameterised solids organize volumes in voxels, fastAerosol utilizes a simpler voxelization algorithm which only considers the average droplet number density and droplet bounding radius in contrast to parameterised solids' algorithm which utilizes the location of the sub-objects in voxelization.

Second, fastAerosol differs from parameterised solids by not needing to load/generate the droplets at the start of the simulation: fastAerosol allows dynamic population of droplets as particles travel through an aerosol (`lazy' population). This allows speedups especially in aerosols where the effective volume (in which particles transport through) is significantly smaller than the total bulk volume or when the number of primaries is low.

Third, the fastAerosol class utilizes the droplet geometry to speed up calculations. The motivation is that, if the distance to the center of a droplet gives knowledge of the distance to the surface $\pm\delta$, then the closest droplet must be in the spherical shell with radial width $R_0\leq R\leq R+2\delta$ for $R_0$ the distance to the closest droplet center.

\section{\texorpdfstring{\MakeLowercase{fast}A\MakeLowercase{erosol}}{fastAerosol} Geometry}

This report presents an aerosol model, `fastAerosol', implemented in GEANT4 and representing an aerosol as a collection of uniformly shaped droplets. Volumes in GEANT4 are described via a hierarchy\cite{bookForApplicationDevelopers}: at the lowest level, a volume is described by its shape in a `solid volume'. Next, the shape is given a material and physical properties in a `logical volume'. Finally, the shape is given a position and rotation in a `physical volume'. Since the presented fastAerosol class is strictly geometrical, this class is on the solid volume level of the hierarchy.

The model consists of a helper class, `fastAerosol', and a solid volume class, `fastAerosolSolid'. The fastAerosol helper class contains the droplet positions, methods to find nearest droplets, and methods to populate droplets; the fastAerosolSolid class contains the GEANT4 requisite solid functions (e.g., DistanceToIn). We refer to the collection of the two classes as `fastAerosol'.

\subsection{Droplet Distribution}
\label{sec:distributionMethods}
The fastAerosol class was designed to be usable with only knowledge of the aerosol's macroscopic properties. While the aerosol's microstate requires description of each droplet's position and rotation, this information may not be available/relevant to the user. To allow the user to be unconcerned of microstate properties, fastAerosol automatically generates droplets in the aerosol based off of an average droplet number density, a position distribution, and a rotation distribution.

By allowing fastAerosol to control the microstate, computational gains can be achieved by lazily populating droplets. To achieve such population, the aerosol is voxelized and droplets are only generated/loaded in voxels which could contain the closest droplet for a given calculation. The voxelization pitch, $p_{\text{grid}}$, is fixed such that there are an average of $N_{\text{droplets/voxel}}$ droplets per voxel (default$=4$; chosen from experimentation to minimize simulation time):
\begin{equation}
    p_{\text{grid}} = \left( \frac{ N_{\text{droplets/voxel}} }{ \langle n_d \rangle } \right)^{1/3}
\end{equation}
for an aerosol with average droplet number density $\langle n_d \rangle$.

These voxels are contained in a vector of length $N_{\text{tot}} = N_x\, N_y\, N_z$ for $N_x$, $N_y$, and $N_z$ the number of voxels in the $x$, $y$, and $z$-directions respectively (appropriately aligned with the bounding box). A voxel with center $(x,y,z)$ is assigned an index $I(x,y,z)$ in this vector given by
\begin{equation}
    \label{eq:index}
    I(x,y,z) = \frac{(x-x_0)+ N_x \left[(y-y_0) + N_y (z-z_0)\right]}{p_{grid}}
\end{equation}
for $(x_0, y_0, z_0)$ the coordinates of the center of the `minimal' voxel. That is, $(x_0,y_0,z_0)$ are defined so that $x_0\leq x$ and $y_0\leq y$ and $z_0\leq z$ for $(x,y,z)$ the coordinates of the center of any voxel in the bounding box's voxelization. We will call $(x-x_0)/p_{grid}$ the voxel $x$-coordinate and similar for $y$ and $z$.

Two values must be determined for each voxel: a random seed, $s(x,y,z)$, and the expected number of droplets per voxel, $\mu(x,y,z)$. First, $s(x,y,z)$ is chosen to be:
\begin{equation}
    s(x,y,z) = I(x,y,z) + s_{global} N_{\text{tot}},
\end{equation}
for $s_{global}$ a random seed for the entire program. Multiplying $s_{global}$ by $N_{\text{tot}}$ guarantees completely new voxel seeds for different integer global seeds.

Second, $\mu(x,y,z)$ is set based off of the spatial average, $\langle n_d \rangle$, and a distribution, $f$. This is done by first calculating a value, $\mu_0$, proportional to the mean:
\begin{equation}
    \mu_0(x,y,z) = \langle n_d \rangle\, \max\left[0, V(x,y,z)\,f(x,y,z)\right]
\end{equation}
where $V(x,y,z)\in [0,1]$ is the fraction of the volume in the voxel that allows placement of a droplet center (inside the aerosol bulk and at least $r$ away from the edge for droplet radius $r$). If $(x,y,z)$ is at least $r+\sqrt{3}p_\text{grid}/2$ from the aerosol surface, then the voxel is fully contained in the aerosol so $V(x,y,z)=1$. Otherwise, this overlap fraction is manually calculated by placing $100$ points with uniform random distribution in the voxel and setting $V(x,y,z)=N_\text{good}/100$ where $N_\text{good}$ is the number of placed points which allow droplet placement.

Since no restrictions are placed on $f$, this $\mu_0(x,y,z)$ is not guaranteed to give the correct $\langle n_d \rangle$, so $\mu(x,y,z)$ must be scaled from $\mu_0(x,y,z)$:
\begin{equation}
    \mu(x,y,z) = \mu_0(x,y,z) \frac{\langle n_d \rangle\, V_B}{\sum_{\text{voxels}} \mu_0(x',y',z')}
\end{equation}
for $V_B$ the volume of the aerosol bulk. This guarantees that $\sum_{\text{voxels}} \mu(x,y,z) = \langle n_d \rangle\, V_B$, as desired.

When performing distance calculations, voxels are lazily populated: if an unpopulated voxel at $(x,y,z)$ could have the closest droplet, a Poisson random number of droplets (mean $\mu(x,y,z)$ and seed $s(x,y,z)$) are placed with centers uniform randomly distributed in the voxel. If a center is to be placed less than $r$ from aerosol boundary or less than $2r+minSpacing$ from any other droplet center (for some user-specified buffer, $minSpacing$), the position is regenerated. If a single center fails placement a user-specified amount of times (default is $100$ attempts), the simulation skips that placement. In the event that the number of skipped centers exceeds a user-specified maximum (default is $1\%$ of the total droplets expected in the aerosol), the simulation ends with an error.

\subsection{Closest Droplets}
GEANT4 distance calculations are sensitive to the droplet with the closest surface, so discussion is focused on determination of said droplet. The fastAerosol class finds the droplet with the closest surface by searching droplet centers for a collection of droplets guaranteed to contain it and then iterating over this collection to find the one with the closest surface. There are two definitions of distance relevant to GEANT4 calculations: absolute distance and distance along a vector, both of which are covered in the following sections.

To generate this collection of candidates for either definition of distance, the relationship between the distance to a droplet's center and the distance to its surface is required. Define $r_\text{in}$ to be the radius of the largest sphere, centered at a droplet, that can be drawn fully inside the droplet; define $r_{\text{droplet}}$ to be the radius of the smallest sphere, centered at the same droplet, that contains the droplet (also called the `bounding radius'). Since droplets are assumed to be uniform in shape/size, these $r_\text{in}$ and $r_\text{droplet}$ hold for all droplets.

For $R_0$ the distance from a query point to the center of the droplet with the closest center, it is then known that the droplet with the closest surface is centered in the shell (centered at the query point) with radius range $R_0<R<R_0+\sigma$ where $\sigma=r_{\text{droplet}}-r_\text{in}$. This $\sigma$ bound is only strictly correct for finding the absolutely closest droplet, but we also use it as an approximation for the closest droplet along a vector.

\subsubsection{Absolute Closest Droplet}
\label{sec:absoluteClosest}
To calculate the distance from a query point $\vec{p}$ to the fastAerosol aerosol, define $R_\text{bulk}$ to be the distance of $\vec{p}$ from the aerosol bulk's surface. Additionally, let $\mathscr{C}$ be the collection of candidate droplets and $R_0$ be the minimum distance from $\vec{p}$ to the center of any droplet in $\mathscr{C}$ (currently $\mathscr{C} = \{\}$ so we treat $R_0=\infty$).

To fill $\mathscr{C}$, define a `voxel radius' $R_{voxel}=\left \lfloor R_\text{bulk}/p_{grid}\right \rfloor$ and search spherical shells of voxels centered at $\vec{p}$ with radius $R\geq R_{voxel}$ (shells generated by the modified mid-point circle algorithm defined in \cite{doi:10.1016/j.jcp.2013.01.035}). Upon finding any droplet, it is added to $\mathscr{C}$ and $R_0$ is updated. The collection is complete when $R>\lceil 0.25+(R_0+\sigma)/p_{grid} \rceil$ or when $R$ exceeds a user defined maximum. The upper limit on $R$ is chosen so that, at this limit, there are no points in the voxelized shell at radius $R+1$ that are closer than $R_0+\sigma$ (upper limit calculated in $2D$ and used as approximation for $3D$).

The absolute closest droplet is then found by iterating over all droplets in $\mathscr{C}$ that have centers closer than $R_0+\sigma$ and returning the droplet with the closest surface distance (return no droplet if $\mathscr{C}=\{\}$).

\subsubsection{Vector Closest Droplet}
To calculate the closest droplet to a particle at point $\vec{p}$ along normalized vector $\vec{v}$, first, define a voxel search width, $w$, as 
\begin{equation}
    w = 2 \left \lceil{\frac{r_{\text{droplet}}}{p_{\text{grid}}}}\right \rceil+1,
\end{equation}
and a temporary position, $\vec{p}\,'$, beginning at $\vec{p}\,' = \vec{p}$ with distance $R_\text{bulk}$ from the aerosol bulk (let $R_\text{bulk}$ automatically update when $\vec{p}\,'$ changes). Additionally, let $\vec{d}$ be the closest droplet found so far and $R_0$ the distance to this droplet's center (treat $R_0=\infty$ at start). The search algorithm, then, is to
\begin{enumerate}
    \item if $R_\text{bulk}=\infty$, return $\vec{d}$; otherwise, if $R_\text{bulk}>p_{\text{grid}}$, update $\vec{p}\,' \to \vec{p}\,' + (R_\text{bulk}-p_\text{grid}) \vec{v}$.
    \item If $|\vec{p}\,'-\vec{p}|$ is greater than a user defined maximum, return $\vec{d}$; otherwise,
    \item search a $w\times w\times w$ grid of voxels centered at the voxel containing $\vec{p}\,'$, calculate the vector distance to the surface of any droplets in this grid and update $\vec{d}$ and $R_0$ appropriately.
    \item If $|\vec{p}\,'-\vec{p}|>R_0+\sigma$, return $\vec{d}$; otherwise,
    \item Find the smallest $n\in\mathbb{N}$ such that $\vec{p}\,' + n\,0.99\,p_\text{\text{grid}}\vec{v}$ lies in a new voxel, define $\vec{\delta}=\delta_x\hat{x} + \delta_y\hat{y} + \delta_z\hat{z}$ to be the difference in voxel positions between the $\vec{p}\,' + n\,0.99\,p_\text{\text{grid}}\vec{v}$ and $\vec{p}\,'$, and update $\vec{p}\,' \to \vec{p}\,' + n\,0.99\,p_\text{\text{grid}}\vec{v}$.
    \item If $R_\text{bulk}>p_{\text{grid}}$, return step $1$; otherwise,
    \item for each non-zero $\delta_\alpha$ for $\alpha\in\{x,y,z\}$, search for droplets the `end caps' $\vec{p}_\text{old}\,'+ (w+1)\delta_\alpha \hat{\alpha} + a\hat{\beta} + b\hat{\gamma}$ for $\hat{\beta},\hat{\gamma}\in\{\hat{x},\hat{y},\hat{z}\}\setminus\hat{\alpha}$ with $\hat{\beta}\neq\hat{\gamma}$ and with $a,b\in[-w,w]\cap\mathbb{Z}$.
    \item If $|\vec{\delta}|>1$, also search for droplets the edges/corners between the end caps.
    \item Update $\vec{d}$ and $R_0$ appropriately. Return to step $4$.
\end{enumerate}

Since the surface distance needs to be calculated to know if a droplet is a candidate (the primary could travel through the droplet's bounding radius but not actually intersect the droplet), the explicit generation of the collection $\mathscr{C}$ and then iteration through $\mathscr{C}$ is unnecessary.

\subsection{Shape Flexibility}
The fastAerosol class accepts any bulk shape in the form of a solid volume object, which is valuable since aerosols often have non-rectangular shapes such as tori\cite{doi:10.3847/0004-637x/818/2/164} and the shapes of atmospheric clouds. Similarly, fastAerosolSolid accepts any droplet shape in form of a solid volume object, with the disclaimer that, while distance calculations to droplets of any shape are accurate, these droplets will be placed as if they were spheres of their bounding radius. Thus fine packing of elongated objects (e.g., sticks) is currently not possible with this package.

By only containing the centers in the fastAerosol helper class, construction of structured droplets is simple: associate multiple fastAerosolSolid objects per fastAerosol helper class. Droplet rotation distributions are allowed.

\subsection{Code Implementation}
The fastAerosol class achieves simpler user implementation by its reliance only on macroscopic properties. While implementation of a parameterised aerosol requires the user to provide droplet positions and rotations at construction, fastAerosol only needs distributions for position and rotation:
\begin{lstlisting}
fastAerosol* aerosol =
    new fastAerosol(``aerosol", bulkShape, boundingR,
                    minSpacing, avgDropNumDens, sigma,
                    positionDistribution);

fastAerosolSolid* solidAerosol =
    new fastAerosolSolid(``aerosolSV", aerosol,
                         dropletShape,
                         rotationDistribution));
\end{lstlisting}
where the first argument to each function is its name and the variables are self-explanatory. The distribution arguments (`positionDistribution' and `rotationDistribution') and `sigma' are optional, taking default values of uniform position/rotation and $\sigma=0$. More detailed implementation may be found in the full source code, available as an advanced example in GEANT4 (as of version 10.7).

\section{Experimental Methods}
\begin{figure*}
	\centering
	\includegraphics[width=\linewidth]{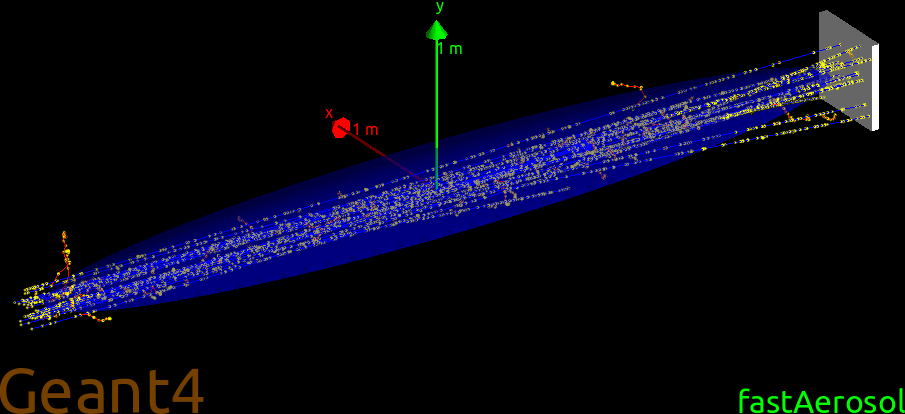}
	\caption{Ellipsoidal aerosol geometry with $50$ MeV protons shot through it from $(0,0,-2612.5)$ mm (with $x$ and $y$ spread $\sigma=16.5$ mm each) along the $+z$-direction.}
	\label{fig:experiment}
\end{figure*}

To test fastAerosol's accuracy and computational efficiency, a fastAerosol aerosol is generated at the origin with a $150\times150\times5000$ mm bounding box (see \cref{fig:experiment} displaying an ellipsoidal aerosol) with varying droplet size, droplet number density, and bulk/droplet shape (droplet size/shape uniform for all droplets in aerosol). Transport is verified by shooting a variable number of $50$ MeV protons along the $z$-axis from a Gaussian distribution centered at $(0,0,-2612.5)$ mm with standard deviation $\sigma=16.5$ mm in the $x$ and $y$ directions. The global seed is allowed to vary across $s_{global}\in[1,5]\cap\mathbb{N}$ for statistics.

The following experiments were performed on GEANT4 10.04 (patch 02) running on an Amazon Web Services (AWS) `c5.24xlarge' computer. We have provided the source code as an advanced example in GEANT4 (as of version 10.7). The data reflects code compiled in release mode and ran in multi-threaded mode.

\subsection{Droplet Distribution}
\label{sec:distribution}
For the droplet distribution tests, droplets with bounding radius $r=1$ mm and number densities of $n=10^{-6+k/5}$ mm$^{-3}$ for $k\in\mathbb{Z}\cap [0,20]$ were populated into a box-shaped aerosol. Droplet placement in the aerosol was performed by individually populating voxels in order of increasing $o(x_i,y_i,z_i)$, defined as 
\begin{equation}
    \label{eq:order}
    o(x_i,y_i,z_i) = z_i + N_z(y_i + N_y x_i)
\end{equation}
for voxel position $(x_i,y_i,z_i)$. This is effectively in order of voxel index (\cref{eq:index}) except with $x$ and $z$ swapped. The distribution of droplets is then saved and analyzed to verify that droplets are roughly uniform in the clouds and thus actually modeling the aerosol of interest.

\subsection{Particle Transport}
Five particle transport experiments were performed to test the following:
\begin{enumerate}
    \item the dependence of particle transport on droplet size,
    \item the dependence of relative computational efficiency between fastAerosol and parameterised aerosols on droplet number density,
    \item the dependence of this relative computational efficiency on primary count,
    \item the dependence on parameterised computational efficiency on its voxelization fineness, and
    \item the dependence of particle transport on bulk/droplet shape.
\end{enumerate}
Four aerosol models are used for these tests: a parameterised model, a fastAerosol model, a fastAerosol model which has droplet positions generated before shooting any primaries (`pre-populated' fastAerosol; as in \cref{sec:distribution}), and a single volume model (`smooth'; density set as average aerosol density). The parameterised aerosol loads droplet positions generated by the pre-populated fastAerosol.

For droplet size investigations, $5000$ protons are shot through parameterised, fastAerosol, pre-populated fastAerosol, and smooth box-shaped aerosols consisting of $n=0.01$ mm$^{-3}$ spherical droplets with radius $0.1\leq r\leq1.50$ mm. The energy deposited into the detector aluminum block is recorded by a $20\times 20\times 1$ scoring grid.

For droplet number density efficiency tests, $5000$ protons are shot through parameterised, fastAerosol, and pre-populated fastAerosol box-shaped aerosols consisting of spherical droplets with radius $r=1$ mm and number density $n=10^{-6+k/5}$ mm$^{-3}$ for $k\in\mathbb{Z}\cap [0,20]$. The total simulation time along with the program's maximum resident set size was measured via `/usr/bin/time'.

For primary count efficiency tests, $1$ to $10000$ protons are shot through parameterised, fastAerosol, and pre-populated fastAerosol box-shaped aerosols consisting of $n=10^{-3}$ mm$^{-3}$ spherical droplets with radius $r=1$ mm. The simulation time draw was measured by `/usr/bin/time'.

For parameterised solid voxelization tests, $5000$ protons are shot through a box-shaped aerosol consisting of $n=10^{-3}$ mm$^{-3}$ spherical droplets with radius $r=1$ mm. The `smartless' parameter, defined as the average number of voxels per geometry object in the parameterised solid, measures the fineness of parameterised solids' voxelization (higher value implies finer voxelization)\cite{bookForApplicationDevelopers}. GEANT4 automatically calculates an `optimal' smartless value and then sets the real smartless value as the minimum of the calculated and user-set values. To test how the voxelization fineness affects simulation time and memory, the user-set smartless parameter is allowed to vary from $2^{-14}$ to $2^5$ and  both simulation time and memory draw were measured by `/usr/bin/time'.

Finally, for droplet shape tests, $5000$ protons are shot through a box-shaped fastAerosol aerosol with $n=0.01$ mm$^{-3}$ droplets shaped like cylinders, boxes, hemispheres, and spheres (all maximally sized to fit in a bounding radius of $r=1$ mm; no rotations applied - aligned along z-axis) and the energy deposited into the aforementioned detector/scoring grid recorded. Similarly, for bulk shape investigations, $5000$ protons are shot through a fastAerosol aerosol containing $n=0.01$ mm$^{-3}$ spherical droplets with radius $r=1$ mm inside a bulk shaped like
\begin{enumerate}
    \item a box ($150\times150\times 5000$ mm),
    \item a cylinder ($r=75$ mm, $\text{length}=5000$ mm; aligned along $z$-axis),
    \item an ellipsoid ($r_\text{minor}=75$ mm, $r_\text{major}=2500$ mm; major radius aligned along $z$-axis), and
    \item a pipe ($r_\text{outer}=75$ mm, $r_\text{inner}=37.5$ mm, $\text{length}=5000$ mm; aligned along $z$-axis).
\end{enumerate}
Again, the energy deposited into the detector/scoring grid was recorded.

\section{Results}
\subsection{Droplet Distribution}
The uniformity of the droplet $x$-position distribution is verified using a one-sample Kolmogorov-Smirnov test with the null hypothesis that the droplet $x$-positions come from a uniform distribution. The distributions from the five seeds were concatenated for better statistics. Under a statistical cutoff of $\alpha=0.05$, the $n=10^{-6}$ mm$^{-3}$ data fails to reject the null hypothesis while the $n=10^{-2}$ mm$^{-3}$ data succeeds in rejecting this hypothesis ($p\approx 44\%$ and $p\approx 2\%$ of drawing a distribution from the null hypothesis more extreme than measured for $n=10^{-6}$ mm$^{-3}$ and $n=10^{-2}$ mm$^{-3}$ respectively). This suggests that, while the $n=10^{-6}$ mm$^{-3}$ data is roughly uniform, the $n=10^{-2}$ mm$^{-3}$ data is not.

\begin{figure}
	\centering
	\includegraphics[width=0.75\linewidth]{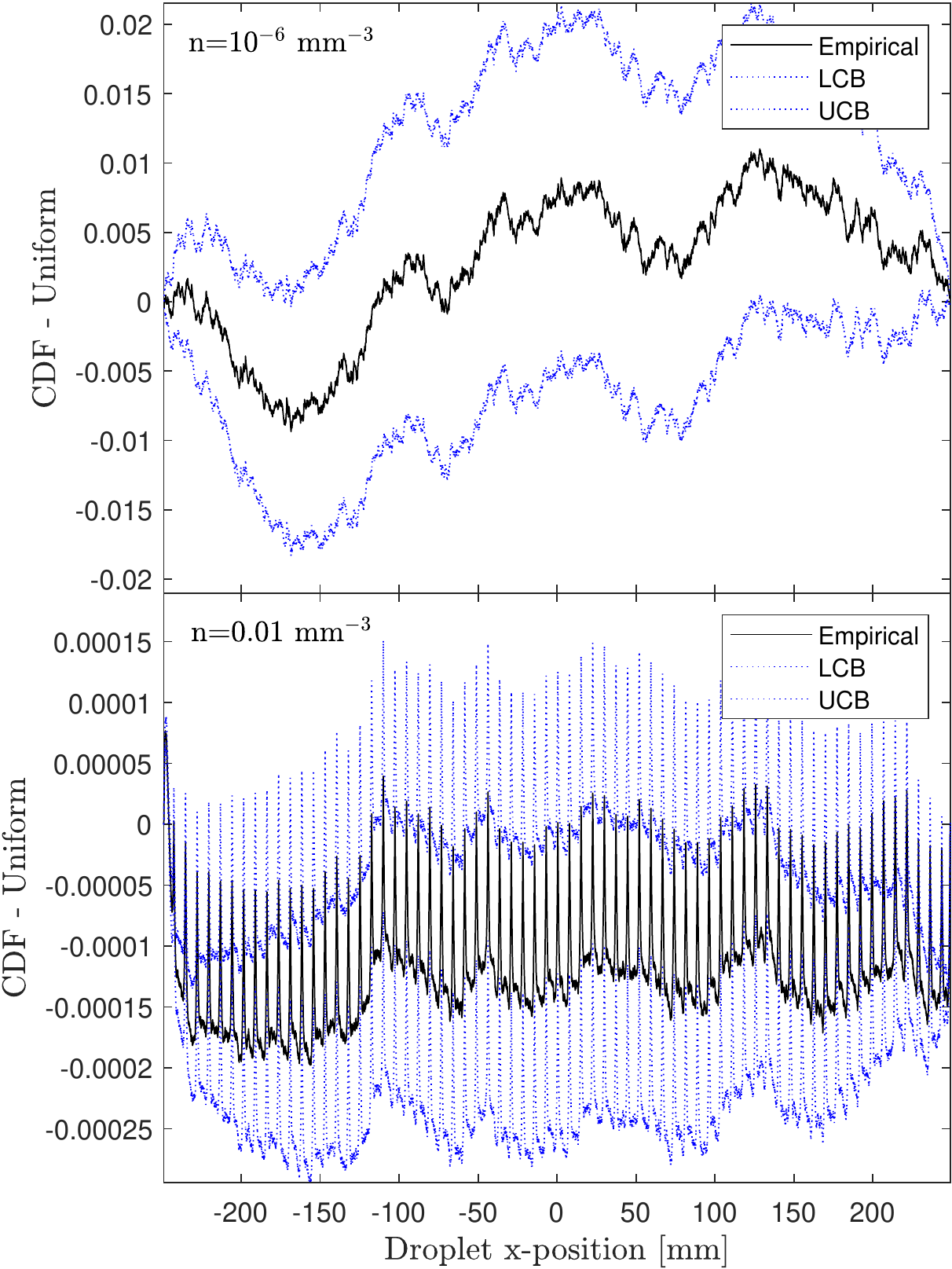}
	\caption{Cumulative distribution plot of droplet $x$-position minus the expected CDF of a uniform distribution. Top: the $n=10^{-6}$ mm$^{-3}$ $x$-position distribution is generally uniform, with the UCB of the observed distribution consistently higher than the uniform expectation and the LCB lower than the uniform expectation. Bottom: the $n=10^{-2}$ mm$^{-3}$ $x$-position distribution is observably non-uniform, with the UCB and LCB both generally smaller than the uniform expectation and a periodic bunching at frequency $\sim7.4$ mm (matching voxel pitch), suggesting a $\sim0.01\%$ heightened probability of droplet placement at voxel boundaries.}
	\label{fig:distribution_plot}
\end{figure}

This failure of uniformity can be visualized in \cref{fig:distribution_plot}, displaying in a solid black line the observed cumulative distribution function (CDF) of droplet $x$-position (other directions similar) minus the expected uniform distribution. The 95\% lower confidence bound (LCB) and upper confidence bound (UCB) are plotted in dotted blue lines, similarly minus the expected uniform distribution. The CDF for $n=10^{-6}$ mm$^{-3}$ data is displayed on top; the CDF for $n=0.01$ mm$^{-3}$ data is displayed on bottom.

For truly uniform data, one would expect the predicted uniform CDF would lay between the observed LCB and UCB, implying that, on \cref{fig:distribution_plot}, the UCB-Uniform curve should be mostly positive while the LCB-Uniform curve mostly negative. This is observed in the $n=10^{-6}$ mm$^{-3}$ data, as expected since our statistical test failed to distinguish this data from a uniform distribution. The $n=10^{-2}$ mm$^{-3}$ data, however, does not show this feature. Namely, the empirical data, LCB, and UCB are all consistently lower than a uniform distribution, except for periodic spikes in the CDF. This indicates that the observed droplet number density is, on average, low, with periodic bunching. The bunching period of $\sim7.4$ mm matches the grid pitch of $7.368$ mm and, suggesting heightened droplet population at voxel boundaries.

Despite this deviation, the overall distribution is still qualitatively uniform: these spikes in probability correspond to an increased likelihood of finding a droplet with $x$-position near a voxel boundary of $\sim 0.01\%$. This level of uniformity is expected to be sufficient for all but the most precise simulations. Regardless, this bunching should be resolved in future work.

\subsection{Droplet Size}
\begin{figure}
	\centering
	\includegraphics[width=0.75\linewidth]{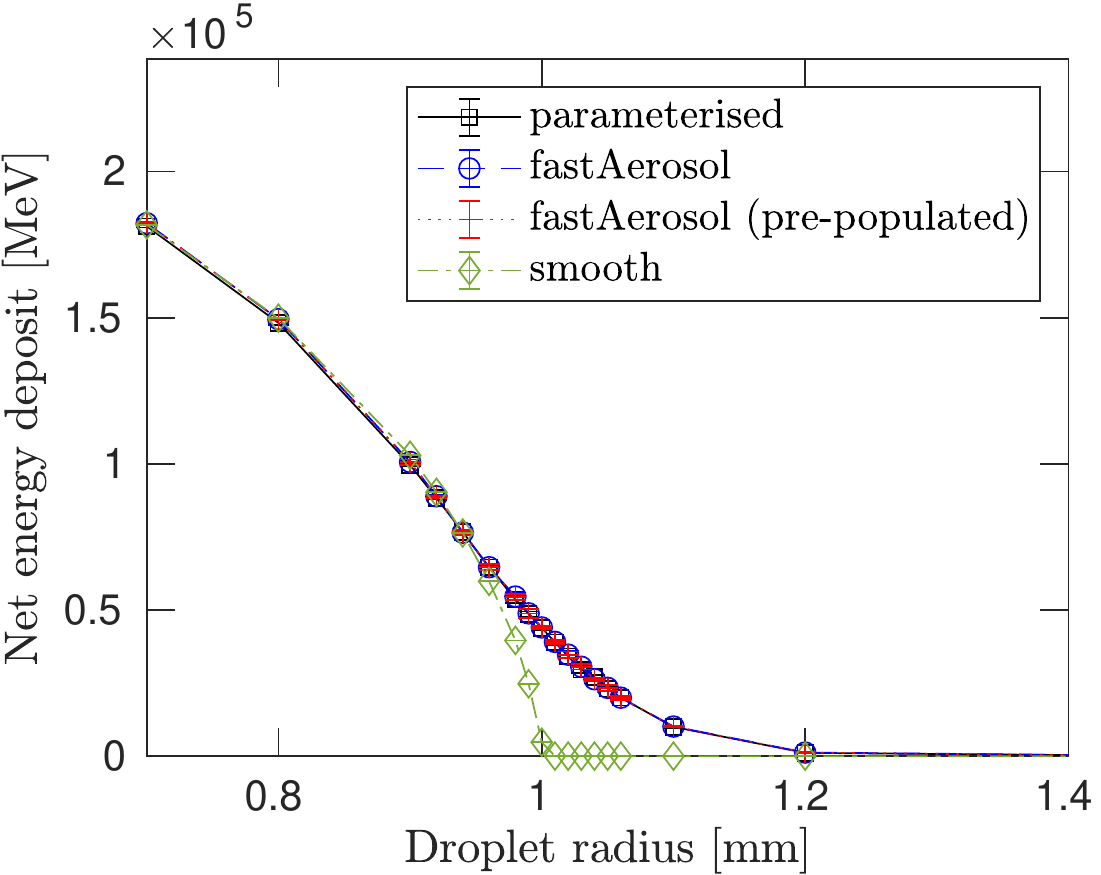}
	\caption{Plot of the energy deposited into the $150\times 150\times 50$ mm aluminum block at $(0,0,2625)$ mm by $5000$ protons ($50$ MeV) after travelling through a box-shaped aerosol ($n=0.01$ mm$^{-3}$). The smooth aerosol displays a drop off at $r=1$ mm which is absent from the other simulations, indicating a difference in underlying simulations.}
	\label{fig:net_plot}
\end{figure}

The energy deposited into the detector after a parameterised, fastAerosol, pre-populated fast-Aerosol, and smooth aerosol is shown in \cref{fig:net_plot} (error bars too small for visualization). The energy deposited after the smooth aerosol drops off approaching $r=1$ mm, a feature absent from all of the other simulations. This most likely indicates that, for `large' droplet radii, the granularity is important for the particle transport, as expected since aerosols with large droplets have a heightened probability for multiple interactions in quick succession, something absent from the single volume model.

Above a droplet radius of $r=0.9$ mm, the parameterised simulations show a statistically equal energy deposit to both the dynamically populated fastAerosol and the pre-populated fastAerosol using a p-value of $0.05$ (tested with Welch's t-test). Below $r=0.9$ mm (only two of the five such data points displayed) there is a statistically significant deviation indicating some difference in the simulations. Despite this, the deviation is on the order of $100$s of MeV, which is only around $1\%$ of the total energy deposited.

This discrepancy can likely be explained by the approximations used for determining $\sigma$ (only strictly correct for absolute distance) and for the approximation setting the upper limit in \cref{sec:absoluteClosest} (limit calculated in $2D$ but used for $3D$ geometry).

\subsection{Number Density Efficiency}
\begin{figure}
    \centering
    \includegraphics[width=0.75\linewidth]{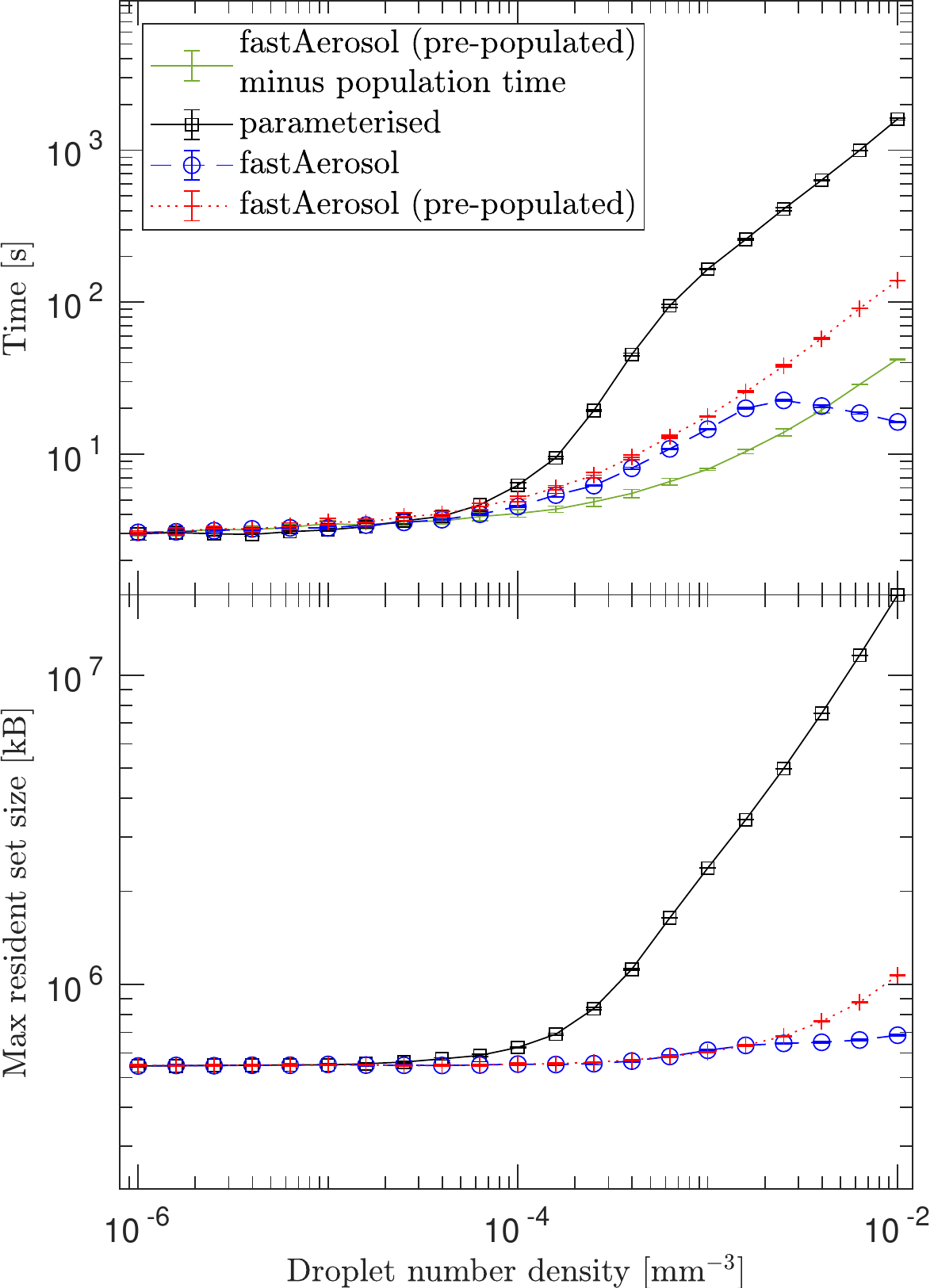}
    \caption{Plots of computational efficiency in simulating the transport of $5000$ protons ($50$ MeV) through a box-shaped aerosol with $r=1$ mm spherical droplets. Top: this plot of simulation time displays that, for all number densities, fastAerosol simulations are quicker than (or comparable to) parameterised simulations with significant speedup at large $n$. The kink in fastAerosol's simulation time at $n\approx 10^{-2.6}$ mm$^{-3}$ is likely due to early stopping of protons, reducing the number of generated droplets. Bottom: this plot of memory usage shows that fastAerosol simulations use less memory than (or comparable to) parameterised simulations for all number densities with a reduction of approximately one order of magnitude at large $n$.}
    \label{fig:efficiency_plot}
\end{figure}

Total simulation time and maximum resident size for the parameterised, fastAerosol, and pre-populated fastAerosol simulations are shown in \cref{fig:efficiency_plot} for varying droplet number density. Note that, for both plots, error bars are typically too small to be visible.

\subsubsection{Timing}
The fastAerosol simulations are significantly faster than (or comparable to) the parameterised simulations for all number densities, with a pronounced speedup beyond number densities of $n=10^{-3}$ mm$^{-3}$. Since fastAerosol simulation times include the generation of the droplet positions while the parameterised simulations do not, the pre-populated fastAerosol simulation minus this population time is also plotted (green solid line).

Both the parameterised and pre-populated fastAerosol simulation times increase roughly polynomially with number density beyond $n=10^{-3}$ mm$^{-3}$ with roughly equal powers. The relatively constant $y$-gap of approximately $1$ order of magnitude indicates that parameterised simulations are roughly $10\times$ slower than fastAerosol simulations.

The dependence of fastAerosol's simulation time on droplet number density appears to largely be due to the population of droplets, as is seen in the large decrease in the pre-populated fastAerosol simulation time when the population time is subtracted. Despite this reduction, the simulation time seems to have the same power dependence on droplet number density, indicating that droplet population is not the sole timing load that is dependent on number density.

The drop-off in the dynamically populated fastAerosol's simulation time at $n\approx 10^{-2.6}$ mm$^{-3}$ is likely due to early termination of particles, reducing the effective aerosol volume and thus reducing the effective droplet count. This both provides evidence that droplet population is a large factor in the simulation load and it shows the value of the dynamically populated fastAerosol variant: it can be extremely quick compared to the pre-populated simulation models when the effective aerosol volume is significantly smaller than the bulk volume or when few primaries are shot. Specifically, at the largest number density, the dynamically populated fastAerosol's simulation time is approximately two orders of magnitude less than the parameterised simulation time.

\subsubsection{Memory}
For all number densities, the fastAerosol simulations use significantly less memory than (or comparable to) the parameterised simulations, peaking at a discrepancy of around one order of magnitude reductions at large number densities. With the current data, it is hard to determine the scaling of fastAerosol (pre-populated and dynamically populated) simulations at large number densities and whether these match the parameterised memory scaling

\subsection{Primary Count Efficiency}
\begin{figure}
    \centering
    \includegraphics[width=0.75\linewidth]{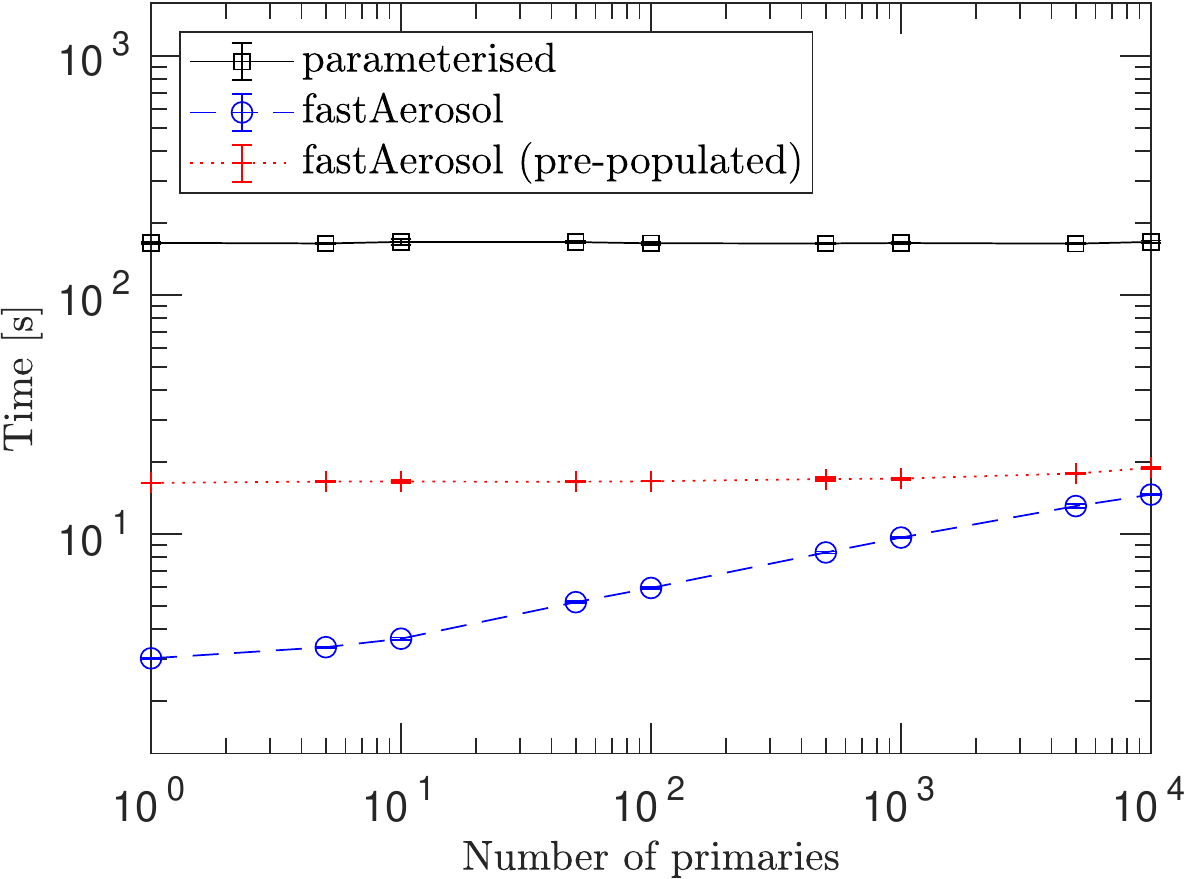}
    \caption{Plots of simulation time after shooting $50$ MeV protons through a box-shaped aerosol ($n=0.01$ mm$^{-3}$; $r=1$ mm) for primary count ranging from $1$ to $10000$. Regardless of this count, the fastAerosol simulations are around $10\times$ (or more) quicker. Both parameterised and the pre-populated fastAerosol simulations appear relatively independent of the primary count while the dynamically populated fastAerosol seems approximately polynomially dependent on this count.}
    \label{fig:efficiency_numPrimaries_plot}
\end{figure}

Total simulation time for the parameterised, fastAerosol, and pre-populated fastAerosol simulations is shown in \cref{fig:efficiency_numPrimaries_plot} for varying primary count (error bars are too small to be visible).

The pre-populated fastAerosol maintains a speedup of approximately a factor of $10$ over parameterised simulations for all primary counts, indicating computational gains due to the simpler voxelization optimization utilized in the fastAerosol class (other speedups expected to change with particle count).

While the pre-populated fastAerosol simulation times are roughly independent of primary count (minor upwards slope not visible in plot), the dynamically populated fastAerosol simulation times are roughly polynomially dependent on primary count. This dependence is most likely due to droplet population since, at small primary counts, the dynamically populated fastAerosol can generate significantly fewer droplets than the pre-populated variant. Since these classes are otherwise identical, it is expected that both fastAerosol variants converge to roughly the same simulation time as the number of primaries becomes large enough for the dynamically populated fastAerosol to populate all voxels.

While not visible in the current plot, there is a positive dependence of simulation time on primary count for the pre-populated fastAerosol simulations and no such dependence for parameterised solids. It would be interesting to see how these dependences extend to extremely high primary counts, such as billions of primaries.

\subsection{Smartless Efficiency}
\begin{figure}
    \centering
    \includegraphics[width=0.75\linewidth]{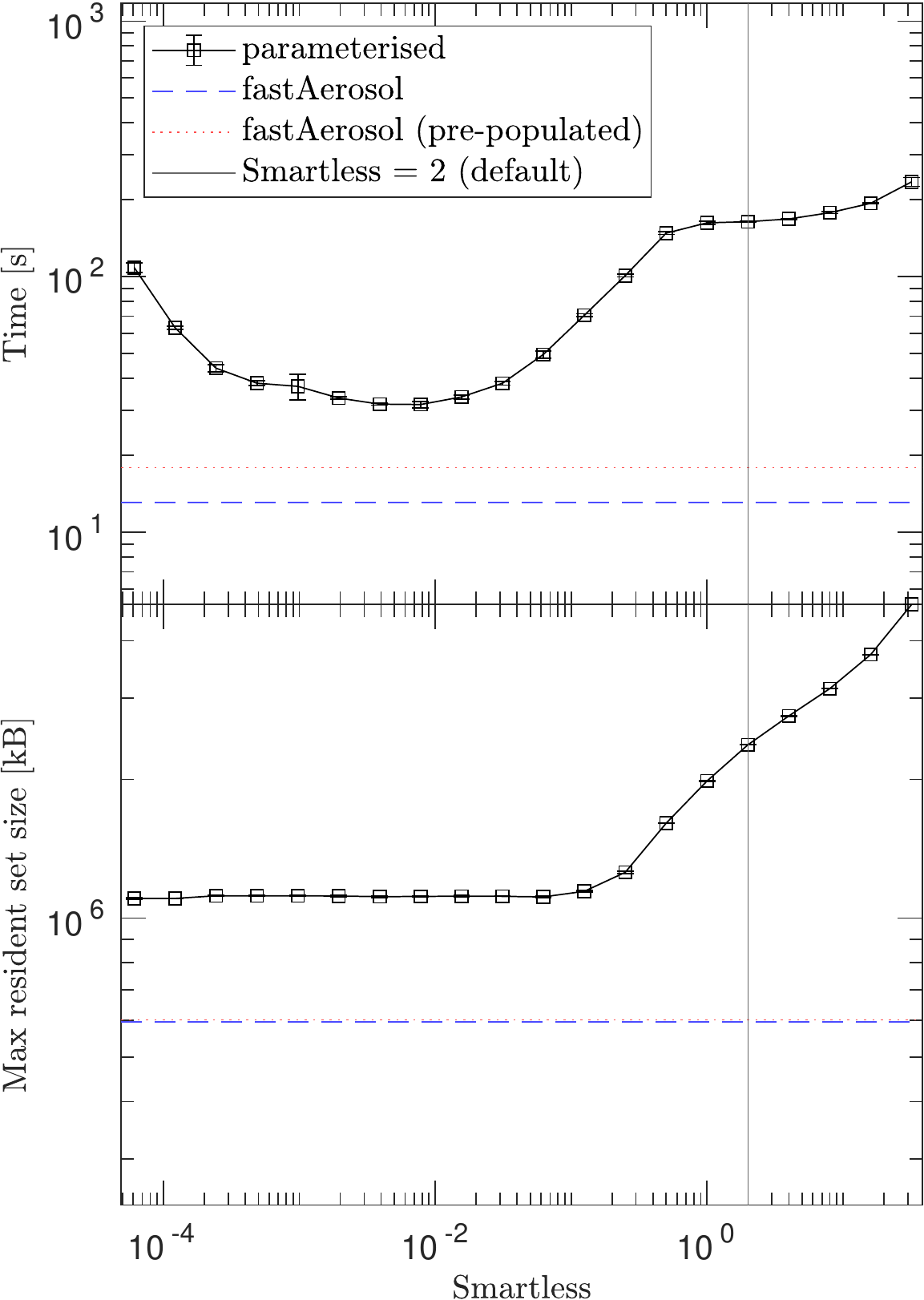}
    \caption{Plots of computational efficiency after shooting $50$ MeV protons through a box-shaped aerosol ($5000$ primaries; $n=0.01$ mm$^{-3}$; $r=1$ mm) for smartless parameter ranging from $2^{-14}$ to $2^5$. It is observed both that the default smartless value does not give a minimum in simulation time/memory and that, even at the true minimum simulation time, fastAerosol is more efficient.}
    \label{fig:efficiency_smartless_plot}
\end{figure}

Total simulation time and maximum resident set size for the parameterised simulations is shown in \cref{fig:efficiency_smartless_plot} for varying `smartless' parameter, defined as the average number of voxels per contained geometry object (default is $2$). The fastAerosol (dynamically populated and pre-populated) average simulation time and memory load are shown as horizontal lines (dashed blue and dotted red, respectively).

Both the memory and simulation time appear dependent on the user-set smartless value, implying that the user-set value is consistently smaller than the calculated value for this aerosol (since the real smartless value is set as the minimum of these two; maximum user-set smartless is $32$). This suggests that the tuning algorithm is overridden by the user-set smartless value for this aerosol and thus an automatically tuned voxelization is not expected, contradicting the user manual's recommendation that no manual tuning is needed\cite{bookForApplicationDevelopers}. This need for additional manual tuning is potentially due to the exotic nature of the simulated geometry: the aerosol contains 12,500,000 droplets, significantly more sub-objects than parameterised typically takes.

The observed dependence of simulation time on smartless is expected: too fine a voxelization and the primary will undergo the overhead of travelling through many empty voxels; too coarse a voxelization and the distance to many droplets that are irrelevant to the primary will be calculated since they will all be lumped into a single voxel. Similarly, the observed memory dependence of parameterised simulations with smartless is as expected: increasing with smartless (since higher smartless implies more voxels) but with a minimum value representing the program's base memory draw.

Even when tuning smartless for minimum simulation time, to a value of approximately $2^{-2}$, both variants of fastAerosol (pre-populated and dynamically populated) maintain a simulation time reduction over the optimized parameterised solids. Similarly, both variants of fastAerosol maintains their more memory reductions over parameterised solids even when tuning smartless for minimum memory usage (even at smartless value $2^{-2}$, approximately where a minimum in simulation time occurs).

A more detailed investigation on tuning the smartless parameter should be done to further verify the computational gains of fastAerosol over parameterised solids for a variety of aerosols, since this test only verified fastAerosol's computational gains over parameterised aerosols for aerosols consisting of $n=0.01$ mm$^{-3}$ spherical droplets with $r=1$ mm. It would be valuable to learn if serious computational gains are allowed for other parameterised geometries by tuning this smartless parameter.

\subsection{Shape Flexibility}
\begin{figure*}
	\centering
	\includegraphics[width=\linewidth]{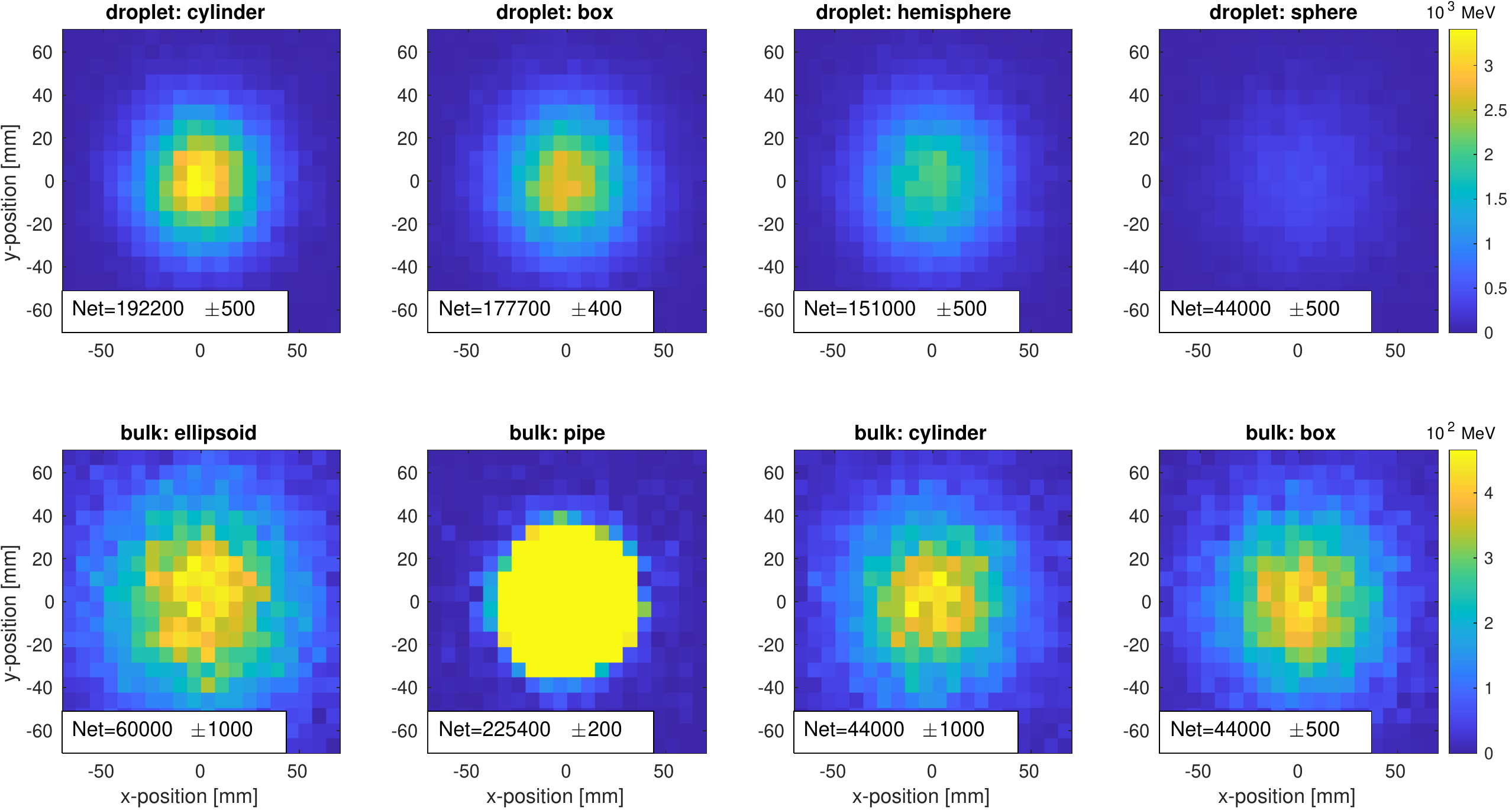}
	\caption{Plots of energy deposited into the $150\times 150\times 50$ mm aluminum block at $(0,0,2625)$ mm, ordered with increasing droplet/bulk filling ratio in the $r=1$ mm bounding radius and in the $150\times 150\times 5000$ mm bounding box moving right, respectively. The energy deposited decreases with increasing filling ratio for the droplet shape tests and for all bulk shape tests except for the pipe, which displays the largest energy deposited by far. This is expected since the pipe presents no barrier to most of the particles.}
	\label{fig:shape_histogram_plot}
\end{figure*}

The energy deposited is recorded after particle transport through a box-shaped aerosol with cylindrical, box-shaped, hemispherical, and spherical droplets (top of \cref{fig:shape_histogram_plot}; chosen to maximally fit in bounding sphere; no rotations applied). Additionally, the energy deposited is recorded after particle transport through a ellipsoid, pipe, cylinder, box-shaped aerosol with spherical droplets (bottom of \cref{fig:shape_histogram_plot}). From left to right, the histograms are placed in order of increasing filling ratio in the bounding sphere/box respectively.

With increasing filling ratio of the droplets in the bounding radius, the net energy deposited is observed to decrease as expected. Similarly, with increasing bulk filling ration in bounding box, the net energy deposited is observed to decrease except for the bulk pipe cloud. The exception of the pipe-shaped cloud is expected since the pipe-shaped cloud presents no barrier to the particles near $(x,y)=(0,0)$, where the majority of particles are shot.

\section{Conclusion}
Presented is a GEANT4 solid, fastAerosolSolid, and a helper class, fastAerosol, which enable accurate and efficient simulations of large aerosols. The fastAerosol classes represent aerosols as a collection of droplets, which we demonstrate is necessary when droplet size becomes large. When droplet number densities become large, the presented fastAerosol model is orders of magnitude more efficient in terms of timing and memory compared to the most efficient droplet-level method of building an aerosol currently included in GEANT4, using parameterised solids, as is seen in \cref{fig:efficiency_plot}.

The computational improvements in fastAerosol are achieved by fastAerosol's simpler voxelization optimization and in the allowed dynamical population of droplets. Additionally, the distance calculations are comparable in time to those performed in parameterised solids as is seen in \cref{fig:efficiency_numPrimaries_plot}: the pre-populated fastAerosol simulation times show a roughly $10\times$ speedup over parameterised simulations regardless of primary count and a higher primary count implies a larger number of distance calculations.

The reduction in memory usage is important for allowing simulations of a wide variety of experiments. The peak observed memory load in simulating particle transport through a parameterised cloud, $18.18\pm 0.01$ GB, is not small when compared to current computer hardware. Simulating the same scenario with the (dynamically populated) fastAerosol class reduces the memory load by a factor of $\sim17\times$ to $1.069\pm0.002$ GB. The parameter space of scenarios that can be simulated, then, is drastically larger when building aerosols with fastAerosol as compared to parameterised solids.

As this will be featured as an advanced example in GEANT4 (available as of version 10.7), it is our hope that the simplicity of implementation of fastAerosol along with the large computational gains will make this class attractive whenever an aerosol needs to be simulated.

\subsection{Future work}
There are some areas for future study to be investigated. First, the periodic droplet bunching apparently at voxel boundaries observed in \cref{fig:distribution_plot} should be studied, maybe by modifying the the order (\cref{eq:order}) in which voxels are filled. Second, the study of the approximations (use of $\sigma$ for vectorized distance and the upper limit on spherical voxel radius in \cref{sec:absoluteClosest}) should be done in hopes of removing the discrepancy in energy deposited after particles transport through aerosols containing droplets with radius $r<0.9$ mm. Third, a more detailed investigation of tuning the smartless parameter for parameterised solids should be done, as that appears to have capacity for large computational gains and lessons beyond the scope of simulating aerosols. It appears possible to disable parameterised voxelization optimization completely, so re-doing tests with this optimization disabled would be interesting in parallel to the smartless investigations to determine if this optimization really does hurt the efficiency.

There are features to add for more sophisticated simulations such as randomized droplet sizes, shapes, and materials. This would require alteration of both the droplet placement and search algorithm, but it seems possible.

Another feature extension that does not seem difficult is to allow a user-specified collision detection function. This would allow accurate placement of non-spherical droplets at the cost of slower simulations. In a similar vein, extending fastAerosol to accept user-specified droplet locations could be valuable, but this would be incompatible with dynamic population.

\subsection{Acknowledgments}
This work built on the GEANT4 source code and examples, so we thank those developers and maintainers. In particular, Dr. Makoto Asai, Dr. Gabriele Cosmo, and  Dr. Susanna Guatelli provided invaluable discussions, recommendations, testing, and assistance with integration with the GEANT4 codebase.





\bibliographystyle{elsarticle-num}
\bibliography{aerosols}







\end{document}